\documentclass[final,3p,times]{elsarticle}
\usepackage{amssymb}

\usepackage[table]{xcolor} % Enables the use of \rowcolor for table rows
\usepackage{multirow}      % Enables the use of \multirow for spanning text across multiple rows
\usepackage{array}         % Provides additional functionality for table formatting
\usepackage{caption}
\usepackage{graphicx}
\usepackage{subcaption}
\usepackage{longtable}
\usepackage{booktabs}
\usepackage{booktabs} 
\usepackage{hyperref}
\usepackage{pdflscape}
\usepackage{mathtools}
\usepackage{array}
\usepackage{float}
\usepackage{setspace}

\begin{document}
\begin{frontmatter}

\title{Complex network analysis of cryptocurrency market during crashes}
\author[inst1]{Kundan Mukhia}
\affiliation[inst1]{organization={Department of Physics, National Institute of Technology},
            %addressline={}, 
            %city={City One},
            postcode={737139}, 
            state={Sikkim},
            country={India}}

\author[inst1]{Anish Rai}
\author[inst1]{S R Luwang}

\author[inst1]{Md Nurujjaman}
\author[inst2]{Sushovan Majhi}
\affiliation[inst2]{organization={The George Washington University},
            addressline={Washington, DC}, 
            %city={City Two},
            postcode={20052}, 
            state={Washington},
            country={USA}}

\author[inst3]{Chittaranjan Hens}
\affiliation[inst3]{organization={International Institute of Information Technology
},
            addressline={Hyderabad}, 
            %city={City Two},
            postcode={500032}, 
            state={Telangana},
            country={India}}    
            
\begin{abstract}

This paper identifies the cryptocurrency market crashes and analyses its dynamics using the complex network. We identify three distinct crashes during 2017-20, and the analysis is carried out by dividing the time series into pre-crash, crash, and post-crash periods. Partial correlation based complex network analysis is carried out to study the crashes. Degree density ($\rho_D$), average path length (\(\bar{l}\)), and average clustering coefficient ($\overline{cc}$) are estimated from these networks. We find that both $\rho_D$ and $\overline{cc}$ are smallest during the pre-crash period, and spike during the crash suggesting the network is dense during a crash. Although \(\rho_D\) and $\overline{cc}$ decrease in the post-crash period, they remain higher than pre-crash levels for the 2017-18 and 2018-19 crashes suggesting a market attempt to return to normalcy. We get \(\bar{l}\) is minimal during the crash period, suggesting a rapid flow of information. A dense network and rapid information flow suggest that during a crash uninformed synchronized panic sell-off happens. However, during the 2019-20 crash, the values of $\rho_D$, $\overline{cc}$, and \(\bar{l}\) did not vary significantly, indicating minimal change in dynamics compared to other crashes. The findings of this study may guide investors in making decisions during market crashes.
 
\end{abstract}

\begin{keyword}
Complex network \sep Hilbert Spectrum \sep Degree density \sep Average path length \sep Average clustering coefficient.
\end{keyword}

\end{frontmatter}
\section{Introduction}
The cryptocurrency market has seen substantial development since its inception in the late 2000s, with a rapid increase in market capitalization and the proliferation of digital currencies~\cite{ji2019dynamic,caporale2018persistence,aste2019cryptocurrency}. As of March 2024, there were more than 23,000 currencies globally even though many of them are thinly traded and the total market capitalization of all these active cryptocurrencies is approximately USD 2.5 trillion~\cite{coinmarketcap}. Cryptocurrencies have garnered significant attention due to various economic factors following the challenges faced by central banks during the 2008 global financial crisis and the 2010-2013 European sovereign debt crisis~\cite{shu2020real}. The decentralized nature, lower transaction costs than traditional fiat currencies, and transaction transparency have attracted a growing number of traders, hence enhancing its popularity~\cite{gupta2022empirical}. However, cryptocurrency as an asset class is still in the nascent stages and as a result, the price of cryptocurrency has had some remarkable fluctuations~\cite{shu2020real}. Further, with the realization of their potential to generate high returns, they have turned into rapidly growing speculative “investment tools”~\cite{antonakakis2019cryptocurrency,baek2015bitcoins,mandaci2022herding}. So, the growth has been coupled with periods characterized by large price fluctuations with increased transaction volume levels~\cite{kim2016predicting, tong2022nonlinear}. Therefore, examining the dynamics and relationship between the return of cryptocurrencies during different price fluctuation periods is of high importance.

Although the popularity of cryptocurrencies has grown, a few studies have been conducted to examine the dynamics and relationship between cryptocurrencies during market crashes. As a result, there is a limited understanding of how the cryptocurrencies interact with one another in terms of return. Prior studies have uncovered the connectedness network among and within different markets that include equities~\cite{fowowe2016dynamic,shahzad2018global,rabindrajit2024high}, bonds~\cite{ahmad2018financial,louzis2015measuring}, currencies~\cite{singh2018dynamic,barunik2017asymmetric}, commodities~\cite{ji2018information,ji2015oil,zhang2020global}, and interest rates~\cite{louzis2015measuring,bech2010topology}. In a study, a time-varying parameters factor augmented vector autoregressive connectedness approach was employed to study the dynamic total connectedness across several cryptocurrencies~\cite{antonakakis2019cryptocurrency}. The dynamic relationships between popular cryptocurrencies and various financial assets were also studied~\cite{corbet2018exploring}. A study on volatility connectedness revealed that volatility spillovers among cryptocurrencies are not necessarily linked to market capitalization, where Bitcoin, despite its significant role, does not dominate the entire market in terms of emitting volatility shocks\cite{yi2018volatility}. Further, spillover effects were quantified across major cryptocurrencies using the connectedness framework of Diebold and Yilmaz~\cite{ji2019dynamic}. Asymmetric multifractal cross-correlation was carried out to examine the cross-correlation behavior between cryptocurrencies and fiat currencies~\cite{fernandes2023asymmetric}. 

Studies related to cryptocurrency market crashes are also very limited. A study on the correlation between Bitcoin and other cryptocurrencies post the major crash in late 2017 to early 2018 was carried out~\cite{yaya2019persistent}. The Bitcoin bubble between December 2017 and December 2018 was also studied, illustrating herding behavior during the Great Crypto Crash and indicating subsequent changes in market quality~\cite{manahov2023great}. During the COVID-19 crash, a comparable investigation into herding behavior was carried out across the four highest-traded cryptocurrencies: Bitcoin, Ethereum, Litecoin, and Ripple~\cite{yarovaya2021effects,susana2020does}. To the best of our knowledge, despite existing studies, no study has yet been carried out that compares the dynamics of different crash periods of cryptocurrency market crashes. Our approach employs complex network analysis based on partial correlation, which provides significant insights into the occurrence of crashes and their relationships with pre-crash and post-crash behaviors. 

In this paper, we analyze cryptocurrency market crashes using complex network analysis. Initially, we used the Hilbert spectrum to identify crashes in 2017-18, 2018-19, and 2019-20. To better understand the nature of these cryptocurrency crashes and their relationships with pre and post-crash periods, we divided each of the three crashes into pre-crash, crash, and post-crash periods. Using partial correlation, we constructed networks for these different market periods. We calculated the degree density ($\rho_D$), average path length (\(\bar{l}\)), and average clustering coefficient ($\overline{cc}$) to analyze the network dynamics during these crash periods. By examining the network changes across different crashes and periods, we aim to uncover the underlying dynamics during crashes.

The rest of the paper is organized as follows:  Sec.~\ref{Methodology} describes the method of analyses and Sec.~\ref{Data analyzed} describes the data employed for the study. It is followed by the results of the study as given in Sec.~\ref{Results}. Finally, Sec.~\ref{Conclusion} gives the concluding remarks.

\section{Method of Analysis}
\label{Methodology}
To study the complex network of cryptocurrencies in different periods of a crash, we applied several methods. Initially, we use the Hilbert spectrum to detect the crash in the cryptocurrency market. It is followed by the evaluation of partial correlation. Next, we employed a threshold based approach to construct a network of cryptocurrencies. Finally, we apply complex network centrality measures to understand the inherent structural properties. Further details on each of these techniques are discussed below.

\subsection{Hilbert Spectrum}
\label{hht1}
The  Empirical mode decomposition (EMD) method is extensively employed for decomposing nonlinear and nonstationary time-series data into distinct intrinsic mode functions (IMFs), each with a specific time scale,  as described in several studies~\cite{huang1998empirical,mahata2020identification,mahata2020time,mahata2021modeling,rai2022sentiment}. A time series must satisfy two conditions to be regarded as an IMF:
\begin{enumerate}

\item The count of extrema and zero crossings must be either identical or differ by a maximum of one.

\item The average of the envelope created by the local maxima and the envelope created by the local minima should be zero.
    
\end{enumerate}

The following steps show the shifting process to acquire IMFs from a time series:

\begin{enumerate}

    \item The upper and lower envelopes of the time series are constructed by connecting their respective maxima and minima using spline fitting.
    
    \item The mean of the envelope is subtracted from the original time series to form a new time series.

    \item  Steps (1) \& (2) are repeated with the new time series until the IMF conditions mentioned above are satisfied. When the conditions are satisfied, the new time series is considered the first IMF.

    \item To identify the second IMF, we replicate the procedure by generating another time series. This series is derived by subtracting the first IMF from the original time series. The decomposition process continues until the monotonic time series is obtained. The original time series can be reconstructed by adding all the IMFs including the monotonic time series.

\end{enumerate}
 To obtain the instantaneous frequency, $\omega$ of an IMF, we perform Hilbert transform defined as
\begin{equation}
    H(t) = \frac{CP}{\pi} \int_{-\infty}^{\infty} \frac{IMF}{t - t'} \, dt,
\end{equation}
where $CP$ is the Cauchy principal. We define  
$\omega$  as
\begin{equation}
     \omega = \frac{d\phi}{dt} \text{, where } \phi_t = \tan^{-1} \left( \frac{H(t)}{IMF} \right).
\end{equation}

Hilbert spectrum (\(H(t, \omega)\)) is the time-frequency distribution and is defined as

\begin{equation}
     H(t, \omega) = \Re \left\{ \sum_{i} K_i(t) e^{j\int \omega(t) \, dt} \right\},
     \label{hh}
\end{equation}

where \( K_i(t) \) is the amplitude. We further estimate the instantaneous energy, \( \overline{E}(t) \) from \( H(t, \omega) \) to identify crashes. \( \overline{E}(t) \) can be estimated as 

\begin{equation}
    \overline{E}(t) = \int_{-\infty}^{\infty} H^2(t,\omega) \, d\omega.
    \label{hht}
\end{equation}

In this analysis, we used Eqn.~\ref{hh} to determine the \(H(t, \omega)\) of the combined IMFs to identify high energy concentration regions in the spectrum, as carried out in these studies~\cite{rai2023detection,mahata2021characteristics}. The high energy concentration regions can be applied to identify sudden changes in a time series~\cite{rai2023detection,mahata2021characteristics}. Hence, this can be applied to identify crashes.

\subsection{Partial Correlation}

The partial correlation is commonly used as a statistical tool to assess the relationship between two variables~\cite{baba2004partial,kenett2010dominating,shapira2009index}. To study the degree of similarity between cryptocurrency price changes, we calculate the daily return of the closing prices. We denote the return of the cryptocurrency, \(p\) at time \(t\) as $R_p(t)$. The average and the standard deviation of \(R_p(t)\) over the interval \([t_i, t_f]\) are given by the following Eqns.~\ref{E} and \ref{S}, respectively:

\begin{equation}
    E[R_p(t)] = \frac{1}{t_f - t_i} \sum_{t=t_i}^{t_f} R_p(t),
    \label{E}
\end{equation}

\begin{equation}
    \sigma[R_p(t)] = \sqrt{\frac{1}{t_f - t_i} \sum_{t=t_i}^{t_f} \left[R_p(t) - E\{R_p(t)\}\right]^2}.
    \label{S}
\end{equation}

Furthermore, the Pearson correlation coefficient~\cite{adler2010quantifying} for the returns $[R_p(t), R_q(t)]$ of two cryptocurrencies $p$ and $q$ is determined as follows:

\begin{equation}
    C_{pq}= \frac{1}{t_f - t_i} \sum_{t=t_i}^{t_f} \left(\frac{R_p(t) - E\{R_p(t)\}}{\sigma\{R_p(t)\}}\right)\left(\frac{R_q(t) - E\{R_q(t)\}}{\sigma\{R_q(t)\}}\right).
\end{equation}

The partial correlation matrix, $C^*_{pq}$ ~\cite{epskamp2018tutorial,stosic2018collective}is then calculated from the inverse of the correlation matrix, $C_{pq}$, where its elements are:

\begin{equation}
    C^*_{pq} = -\frac{C^{-1}_{pq}}{\sqrt{C^{-1}_{pp} C^{-1}_{qq}}},
\end{equation}

The value of \( C^*_{pq} \) ranges from \(-1\) to \(1\), where \( C^*_{pq} = 1 \) indicates that cryptocurrencies \( p \) and \( q \) are strongly correlated, \( C^*_{pq} = -1 \) signifies they are strongly anti-correlated, and \( C^*_{pq} = 0 \) means they are uncorrelated.

\subsection{Network construction}
\label{Threshold Method}

The cryptocurrency network is constructed using the threshold method to simplify the analysis. It reduces the complexity and preserves the most significant relevant features~\cite{xu2017topological}. This method represents cryptocurrencies as interconnected nodes, where an undirected edge connects the nodes \(m\) and \(n\) if the absolute value of \(C^*_{pq}\) meets or exceeds a fixed threshold (\(\theta\)), with \(0 \leq \theta \leq 1\). The threshold method retains all crucial data within the network. The \(\theta\)
used in this study was determined using the percentile method~\cite{xu2017topological}.

A network is defined as, \(G = (V, E, f)\), comprising nodes \(V\), edges \(E\), and a function \(f\) that maps edges to node pairs, $m$ and $n$. In simpler terms, a network without self-loops is represented as \(G = (V, E)\). In a cryptocurrency network, vertices represent individual cryptocurrencies, while edges are established based on the partial correlation coefficient \(C^*_{pq}\) and a threshold \(\theta\). The set of edges \(E\) is defined as~\cite{huang2009network}:

\begin{equation}
E = \begin{cases}
e_{mn} = 1, & \text{if } m \neq n  \text{ and } C^*_{pq} \geq \theta, \\
e_{mn} = 0, & \text{if } m = n.
\end{cases}
\end{equation}

\subsection{Topological features of the network}

\subsubsection{Degree Density}
\label{Degree Density}

Degree density ($\rho_D$), reflecting the average degree or connectivity of nodes, is crucial for understanding the compactness of a network. It is calculated as the ratio of the actual number of edges to the maximum possible number of edges in the network~\cite{nobi2014correlation,rakib2021structure}, given by:

\begin{equation}
\rho_D = \frac{\sum_{k=1}^{N} C_D(k)}{N(N-1)}
\end{equation}

Here, $C_D(k)$ denotes the degree centrality of node $k$ which is defined as the count of links that a node has with other nodes within the same network~\cite{lu2016vital,newman2018networks}. It quantifies the number of direct connections a node has, indicating its interaction level with the network~\cite{battiston2020networks,moghadam2019complex,papana2017financial}. Mathematically, the degree centrality for node \(k\) is expressed as:

\begin{equation}
 C_D(k) = \sum_{i=1}^{n} \alpha(i, k)
\end{equation}

where \(n\) represents the total number of nodes, and \(\alpha(i, k)\) is \(1\) if there is a connection between node \(i\) and node \(k\), and \(0\) otherwise.

\subsubsection{Average path length}

The average path length ($\bar{l}$) is calculated as the average of all shortest path lengths between any pair of nodes~\cite{watts1998collective,jiang2004topological,liu2008complexity}. The shortest path length between two nodes represents the minimum number of connections required to travel from one node to another~\cite{papana2017financial,porta2006network}. The shorter path length indicates that the information diffuses more quickly within the network~\cite{papana2017financial,porta2006network}.  Mathematically, the average path length, $\bar{l}$ is written as, 

\begin{equation}
\bar{l} = \frac{2}{N(N - 1)} \sum_{\substack{m,n \\ m < n}} l_{mn}
\end{equation}

Here, \(l_{mn}\) denotes the shortest distance between nodes \(m\) and \(n\), and \(N\) is the network's total node count.

\subsubsection{Clustering Coefficient}

The clustering coefficient ($cc_j$) quantifies the connectivity level among a vertex's neighbors. It is calculated by the ratio of the actual number of edges linking a vertex's neighbors to the total number of potential edges between them~\cite{watts1998collective,jiang2004topological,porta2006network,liu2008complexity}. The formula used to calculate the clustering coefficient of a given node \(j\) is as follows:

\begin{equation}
cc_j = \frac{2u_j}{v_i(v_j - 1)},
\end{equation}

where \(v_j\) represents the count of neighbors surrounding node \(j\), and \(u_j\) indicates the total edges among those neighbors.

We calculate the average clustering coefficient ($\overline{cc}$)  of the cryptocurrency network by taking the mean of the clustering coefficients of all nodes in the network. The $\overline{cc}$ is given by the formula:

\begin{equation}
    \overline{cc} = \frac{1}{n} \sum_{j=1}^n cc_j
\end{equation}

where \( n \) represents the total number of nodes in the network, and \( cc_j \) is the clustering coefficient for node \( j \).

\section{Data Description: Different cryptocurrencies}
\label{Data analyzed}

To study the network dynamics during various cryptocurrency market crash periods, we utilized daily closing price data obtained from~\cite{coinmarketcap} and~\cite{yahoofinance}. Cryptocurrencies were chosen based on their market capitalization at the time of the crash. The number of cryptocurrencies included in our analysis varies across different crash periods due to irregularities and missing data. Specifically, we analyzed 49 cryptocurrencies for the 2017-18 crash, 35 cryptocurrencies for the 2018-19 crash, and 46 cryptocurrencies for the 2019-20 crash.

\section{Results}
\label{Results}
In Subsec.~\ref{Hilbert Spectrum}, we identify three different crashes in the cryptocurrency market, and Subsec.~\ref{ND} and Subsec.~\ref{NDA} present the results of a partial correlation and complex network analysis of cryptocurrencies in the pre-crash, crash, and post-crash periods of the 2017-18, 2018-19, and 2019-20 crashes respectively.

\subsection{ Crash Identification using Hilbert Spectrum}
\label{Hilbert Spectrum}

\begin{figure}[ht]
    \centering
    \includegraphics[width=8.5cm]{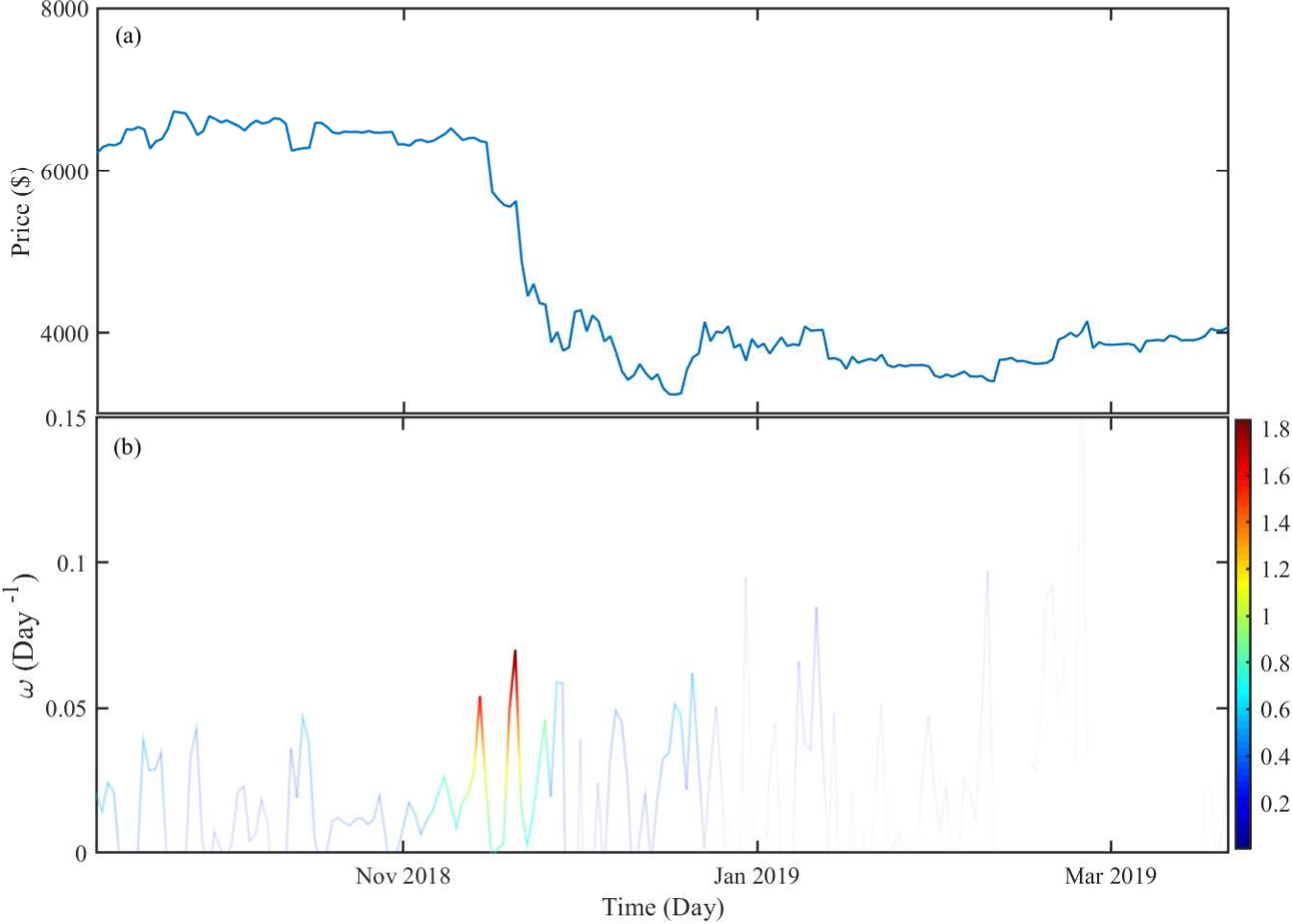}
    \caption{Plot (a) \& (b) represent the daily closing price of BTC-USD from September 2018 to February 2019 and the Hilbert spectrum of the combination of IMFs, respectively.}
    \label{2018-19}
\end{figure}

We estimated the Hilbert spectrum [\(H(t, \omega)\)] of the time series by combining all the intrinsic mode functions (IMFs) to identify a crash in the cryptocurrency market as described in Subsec.~\ref{hht1}. Fig.~\ref{2018-19} (a) and Fig.~\ref{2018-19} (b) represent the daily closing price plot of the  Bitcoin (BTC-USD) from September 2018 to February 2019, and its \(H(t, \omega)\) respectively. The sudden price changes are identified by \(H(t, \omega)\) as shown by the reddish region in the spectrum, indicating the maximum energy concentration, and hence the sharp change in the energy spectrum identifies the crash. Similarly, we have identified crashes during the year 2017-18 and 2019-20 using the \(H(t, \omega)\). The duration of the crash period is taken as the interval from the day with the highest closing price before the crash to the day with the lowest closing price after the crash. 

%Using the \(H(t, \omega)\), we divided the crashes into pre-crash, crash, and post-crash periods and constructed a network for each period.

\begin{table}[ht]
\centering
\caption{Durations of pre-crash, crash, and post-crash periods during the year 2017-18, 2018-19 and 2019-20.}
\begin{tabular}{|>{\centering\arraybackslash}m{2cm}|>{\centering\arraybackslash}m{3cm}|>{\centering\arraybackslash}m{3cm}|>{\centering\arraybackslash}m{3cm}|}
\hline
Crash period & Pre-crash  &  Crash  & Post-crash \\ 
\hline
2017-18 & 01.09.2017 - 16.12.2017 &  17.12.2017 - 05.02.2018 &  06.02.2017 - 20.05.2018 \\ 
\hline
2018-19  &  01.08.2018 - 12.11.2018 &  13.11.2017 - 15.12.2018 &  16.12.2018 - 25.02.2019 \\ 
\hline
2019-20  &  01.11.2019 - 12.02.2020 &  13.02.2020 - 12.03.2020 &  13.03.2020 - 25.06.2020 \\ 
\hline
\end{tabular}
\label{tab:DT}
\end{table}

The periods before and after the crash are considered as the pre-crash and post-crash periods, respectively. For both the pre-crash and post-crash periods, we selected a timeframe of 3.5 months. This selection ensures that the analysis remains focused on the critical transitions associated with the crash. The consistency of results across various durations supports this choice, as extending the period to up to 5 months for both pre-crash and post-crash periods gives similar outcomes. Hence, we have identified the crashes using \(H(t, \omega)\) and divided the crashes into pre-crash, crash, and post-crash periods. Table~\ref{tab:DT} presents the durations of these periods. In the subsequent subsections, we will estimate the partial correlations between the cryptocurrencies and construct the network during these periods of the three crashes. We have omitted the 2022 cryptocurrency crash from our study due to its extended duration, the cessation of operations of some cryptocurrencies used in our analysis, and the emergence of thousands of new cryptocurrencies in 2022. Therefore, a separate analysis of the 2022 crash is required. 

\subsection{Partial correlation of the crashes}
\label{ND}

\begin{figure}[H]
     \centering
     \begin{subfigure}[b]{0.33\textwidth}
         \centering
         \includegraphics[width=\textwidth]{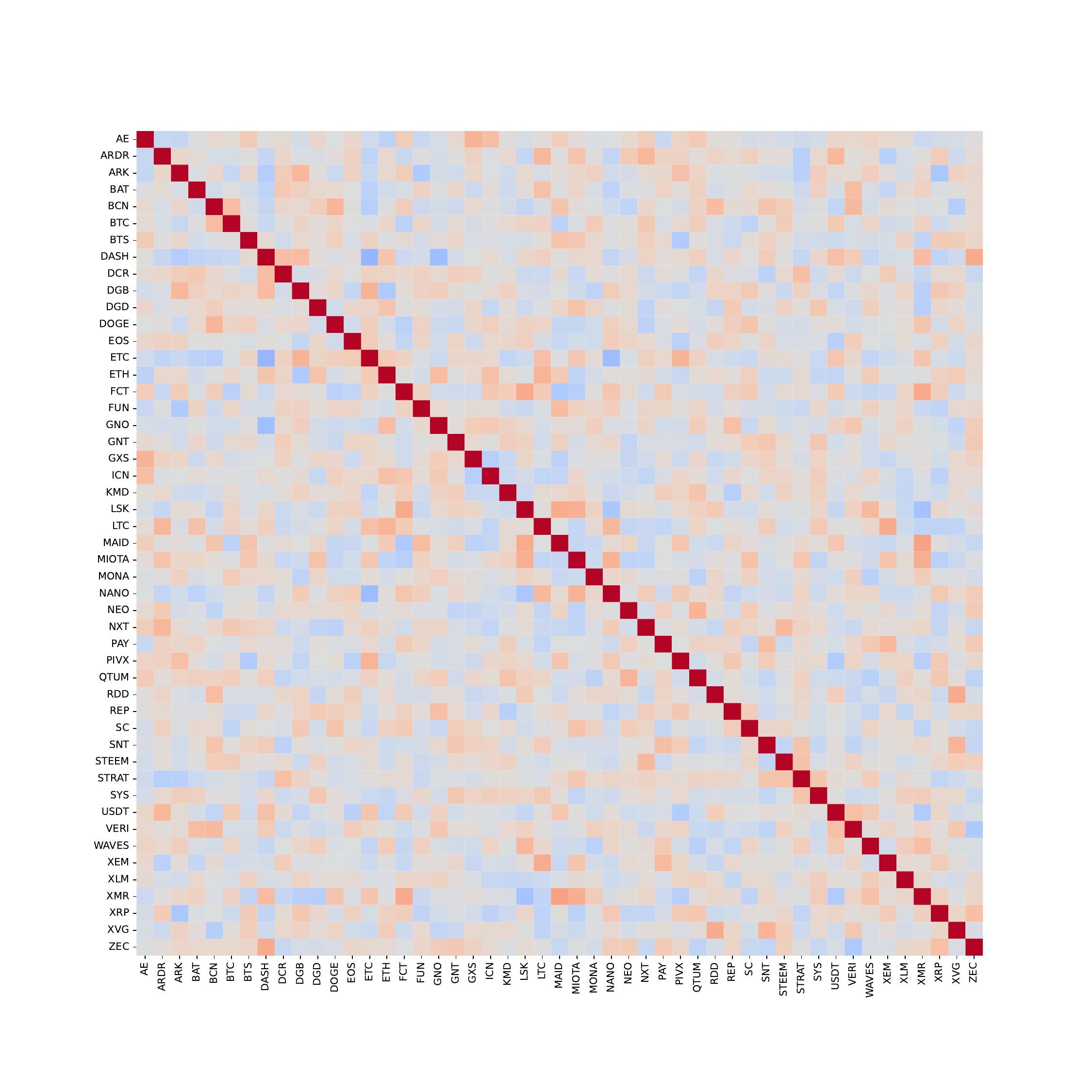}
         \caption{Pre-crash (2017-18)}
         %\label{before17}
     \end{subfigure}
     \hfill
     \begin{subfigure}[b]{0.32\textwidth}
         \centering
         \includegraphics[ width=\textwidth]{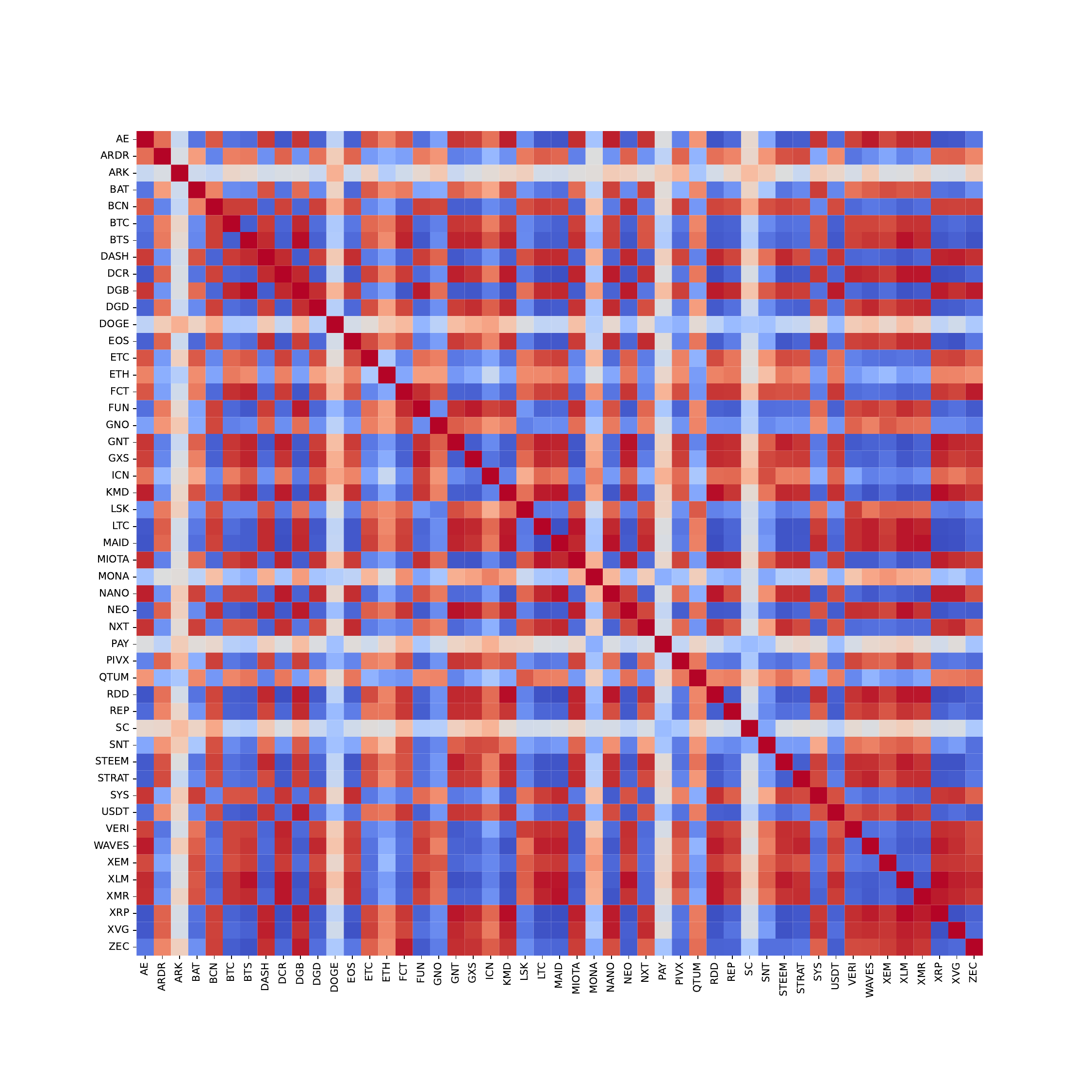}
         \caption{Crash (2017-18)}
         %\label{during17}
     \end{subfigure}
     \hfill
     \begin{subfigure}[b]{0.33\textwidth}
         \centering
         \includegraphics[ width=\textwidth]{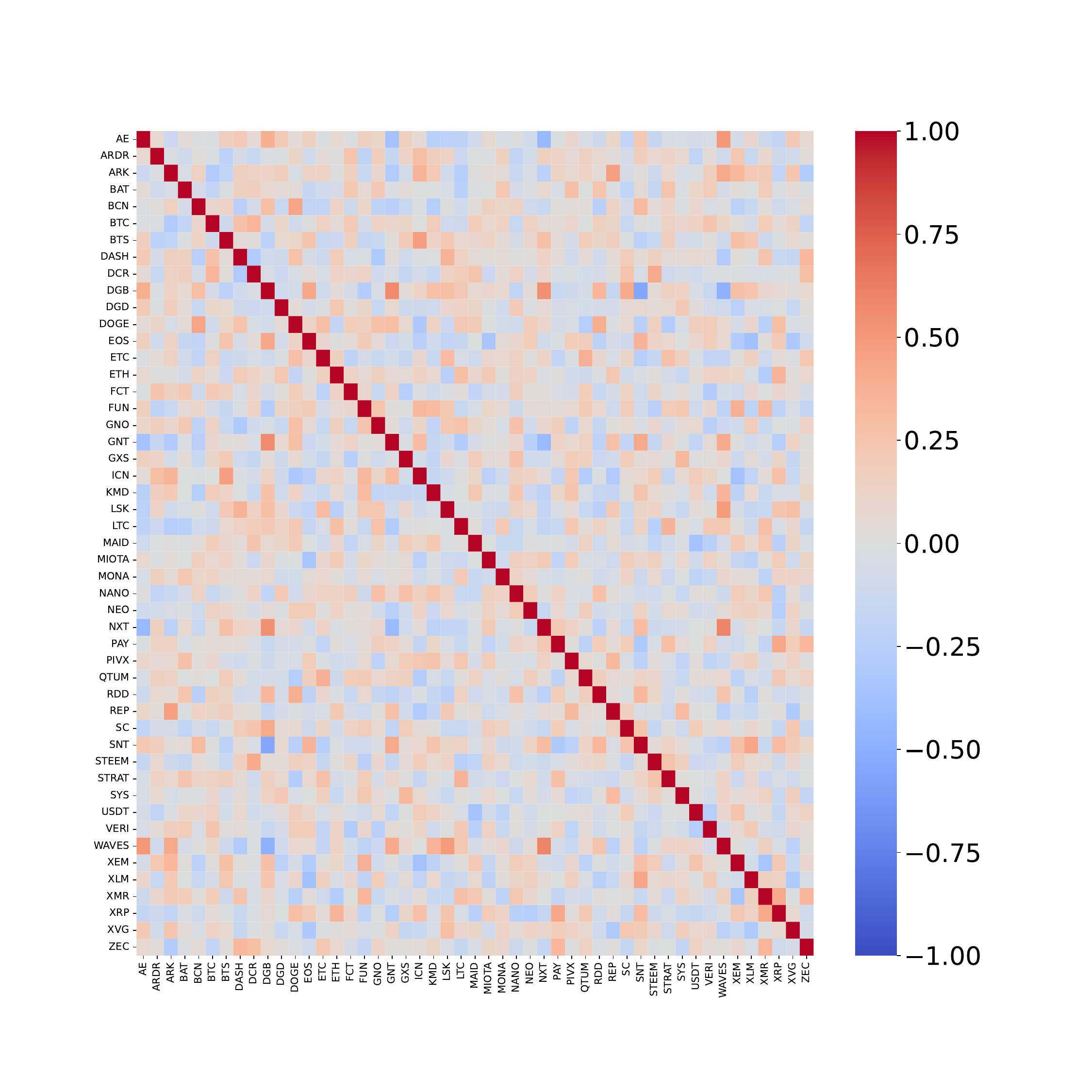}
         \caption{Post-crash (2017-18)}
         %\label{after17}
     \end{subfigure}
      \caption{The heatmap plot represents the partial correlation matrices among different cryptocurrencies. (a), (b) and (c) represent pre-crash, crash, and post-crash periods for the 2017-18 crash.}
        \label{fig:2}
\end{figure}

Figs.~\ref{fig:2}(a), \ref{fig:2}(b), and \ref{fig:2}(c) show the heatmap of the partial correlation among various cryptocurrencies during the pre-crash, crash, and post-crash period of the 2017-18 crash, respectively. In these figures, the color spectrum signifies the strength of the correlation. The intensity of the correlations corresponds to the depth of color, with darker red indicating stronger correlations. From these figures, we observe that the heatmap for the crash period is the most intense than the pre-crash and post-crash periods. This shows that during the crash period, the cryptocurrencies are more correlated. Figs.~\ref{fig:2}(a) and \ref{fig:2}(c) show the correlation is less in the pre-crash and post-crash periods, respectively.  The mean values of the partial correlations during pre-crash, crash, and post-crash of 2017-18 are 0.290, 0.769, and 0.299, respectively.

\begin{figure}[ht]
     \centering
\begin{subfigure}[b]{0.33\textwidth}
         \centering
         \includegraphics[ width=\textwidth]{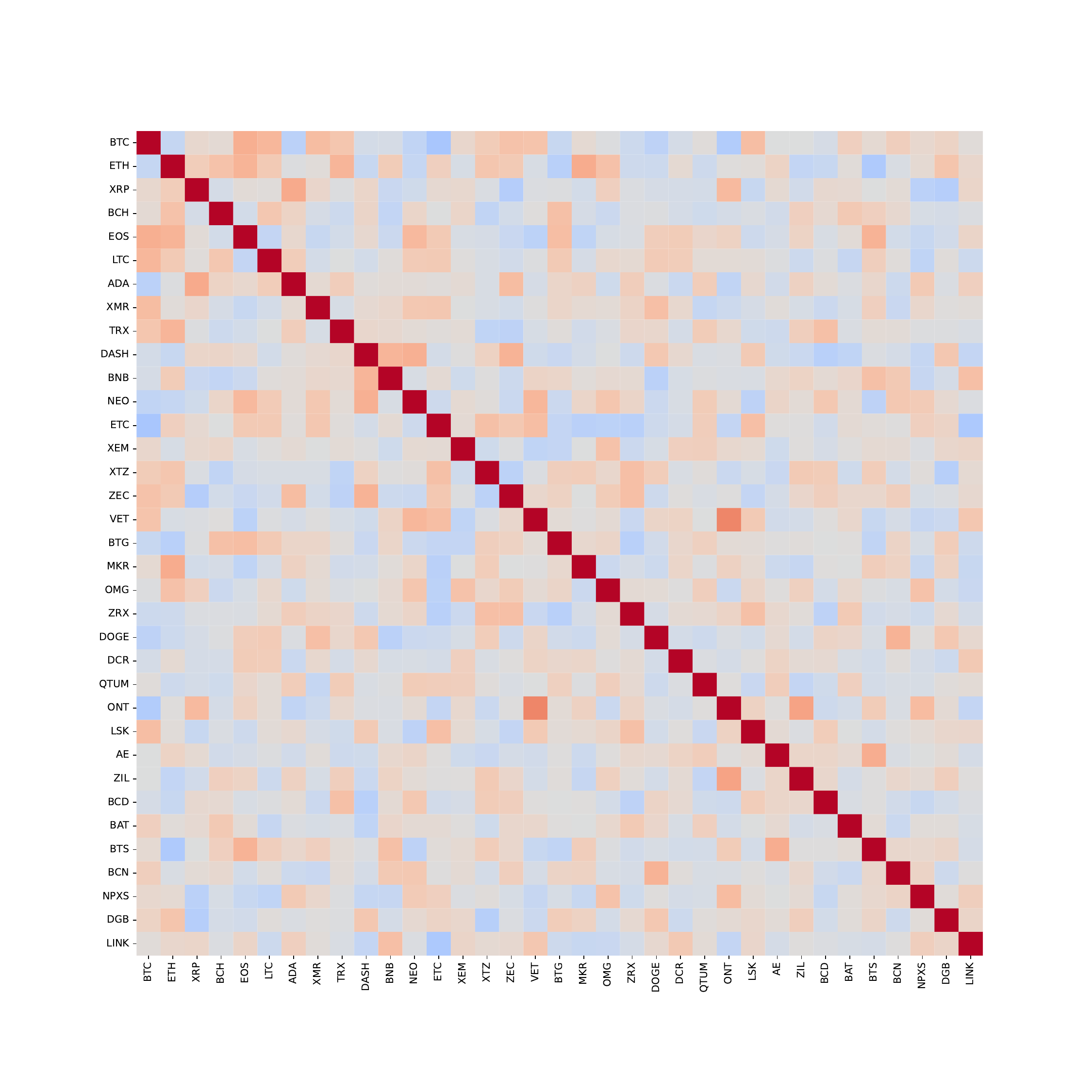}
         \caption{Pre-crash (2018-19)}
         %\label{before18}
     \end{subfigure}
     \hfill
     \begin{subfigure}[b]{0.32\textwidth}
         \centering
         \includegraphics[ width=\textwidth]{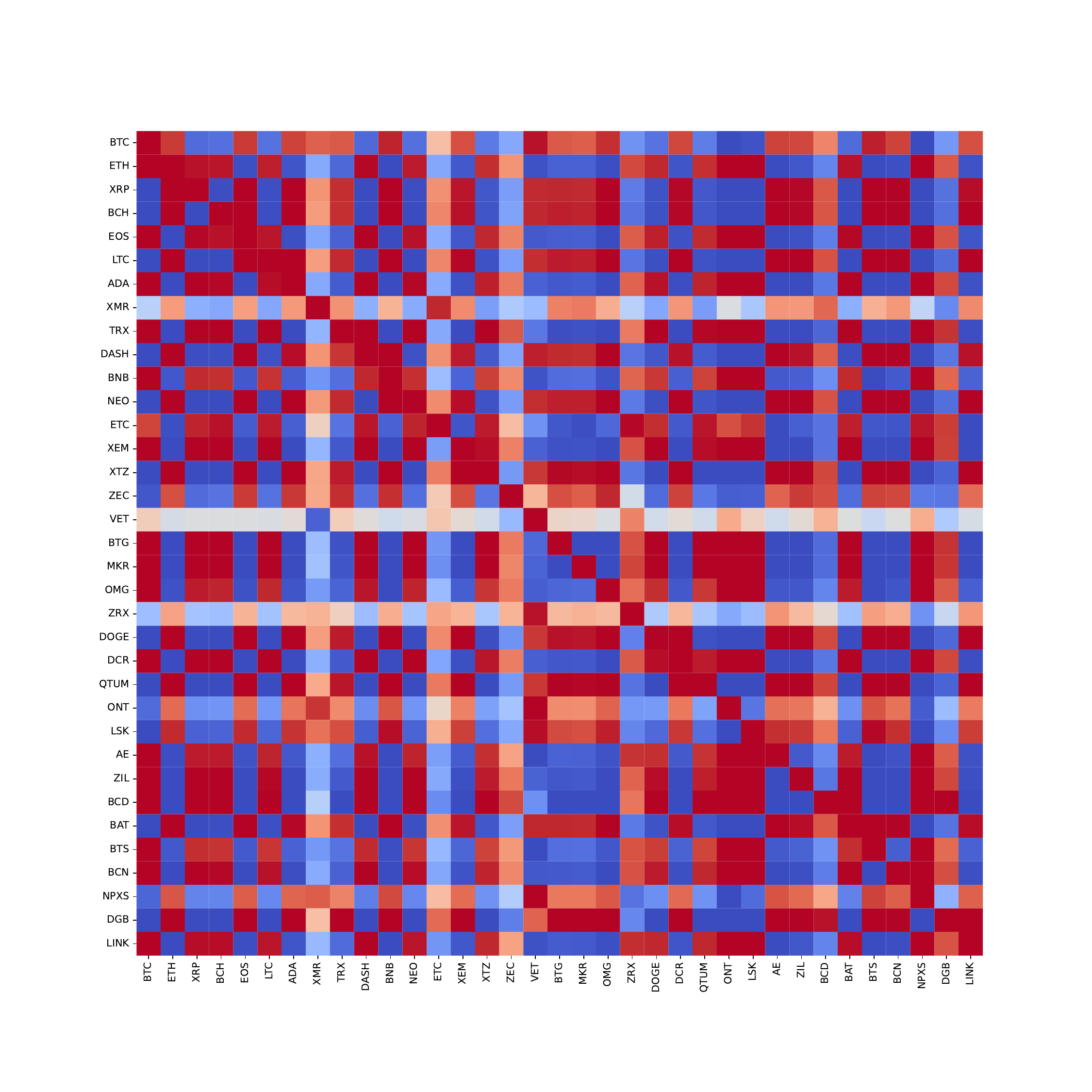}
         \caption{Crash (2018-19)}
         %\label{during18}
     \end{subfigure}
     \hfill
 \begin{subfigure}[b]{0.33\textwidth}
         \centering
         \includegraphics[ width=\textwidth]{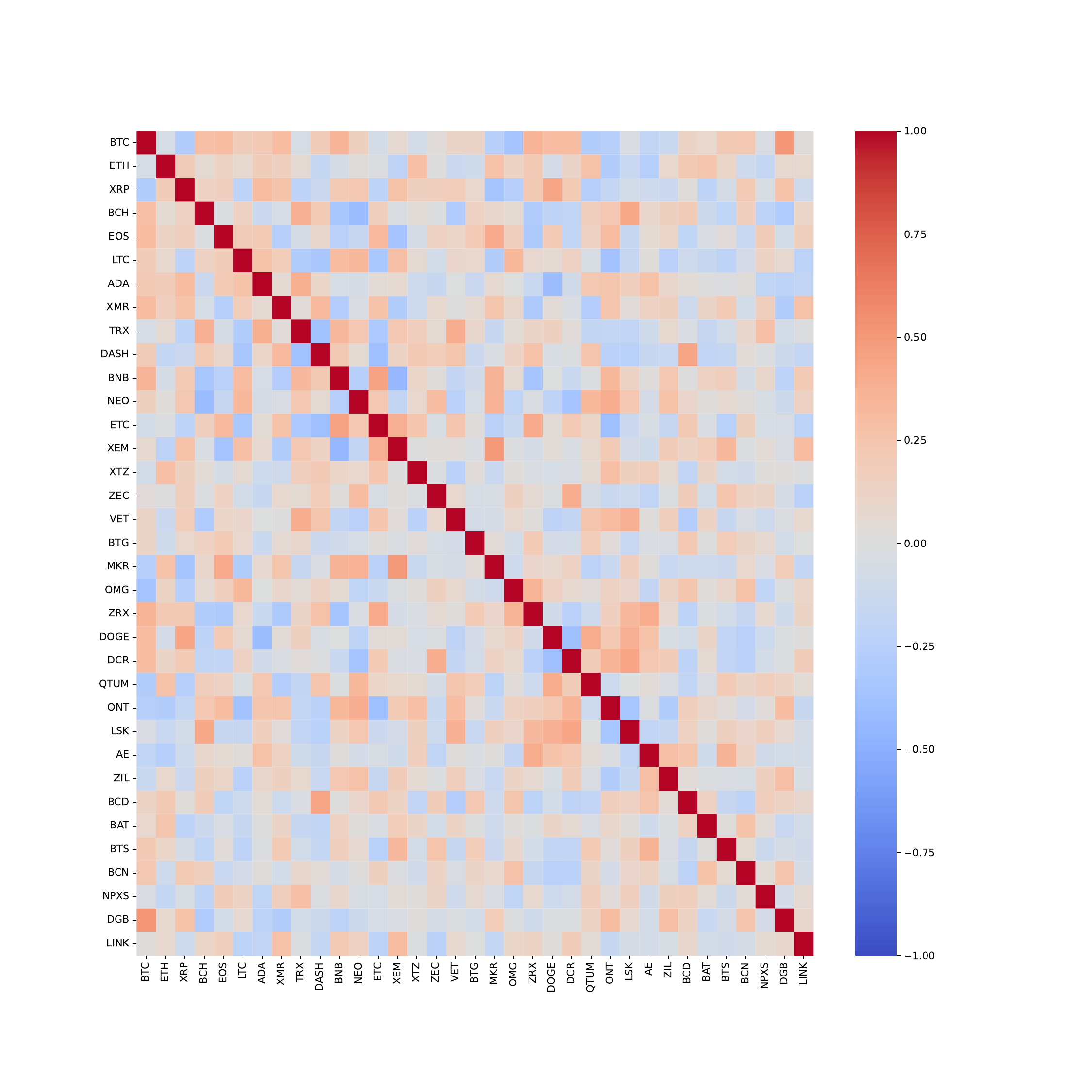}
         \caption{Post-crash (2018-19)}
         %\label{after18}
      \end{subfigure}   
\begin{subfigure}[b]{0.33\textwidth}
         \centering
         \includegraphics[ width=\textwidth]{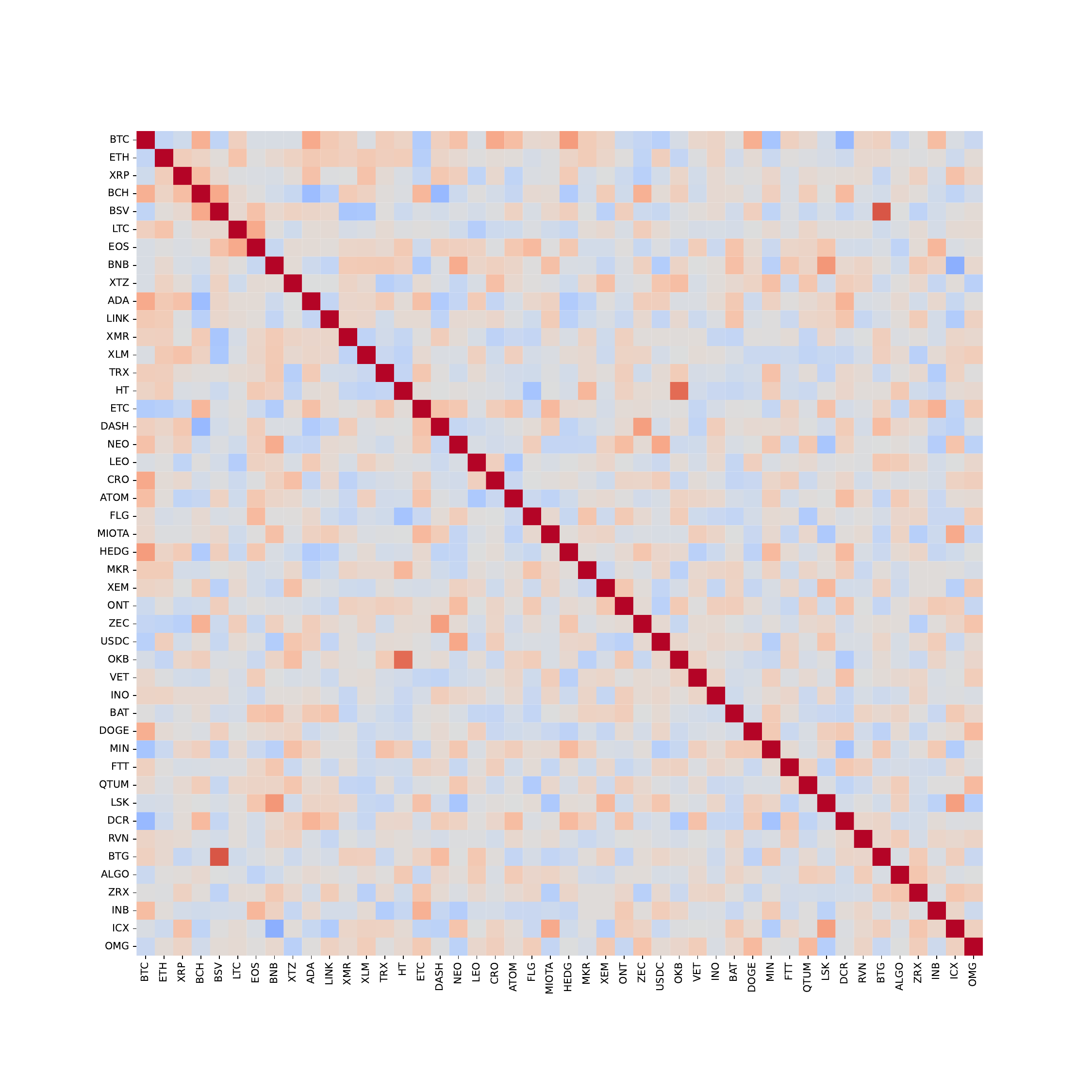}
         \caption{Pre-crash (2019-20)}
         %\label{before19}
     \end{subfigure}
     \hfill
     \begin{subfigure}[b]{0.32\textwidth}
         \centering
         \includegraphics[ width=\textwidth]{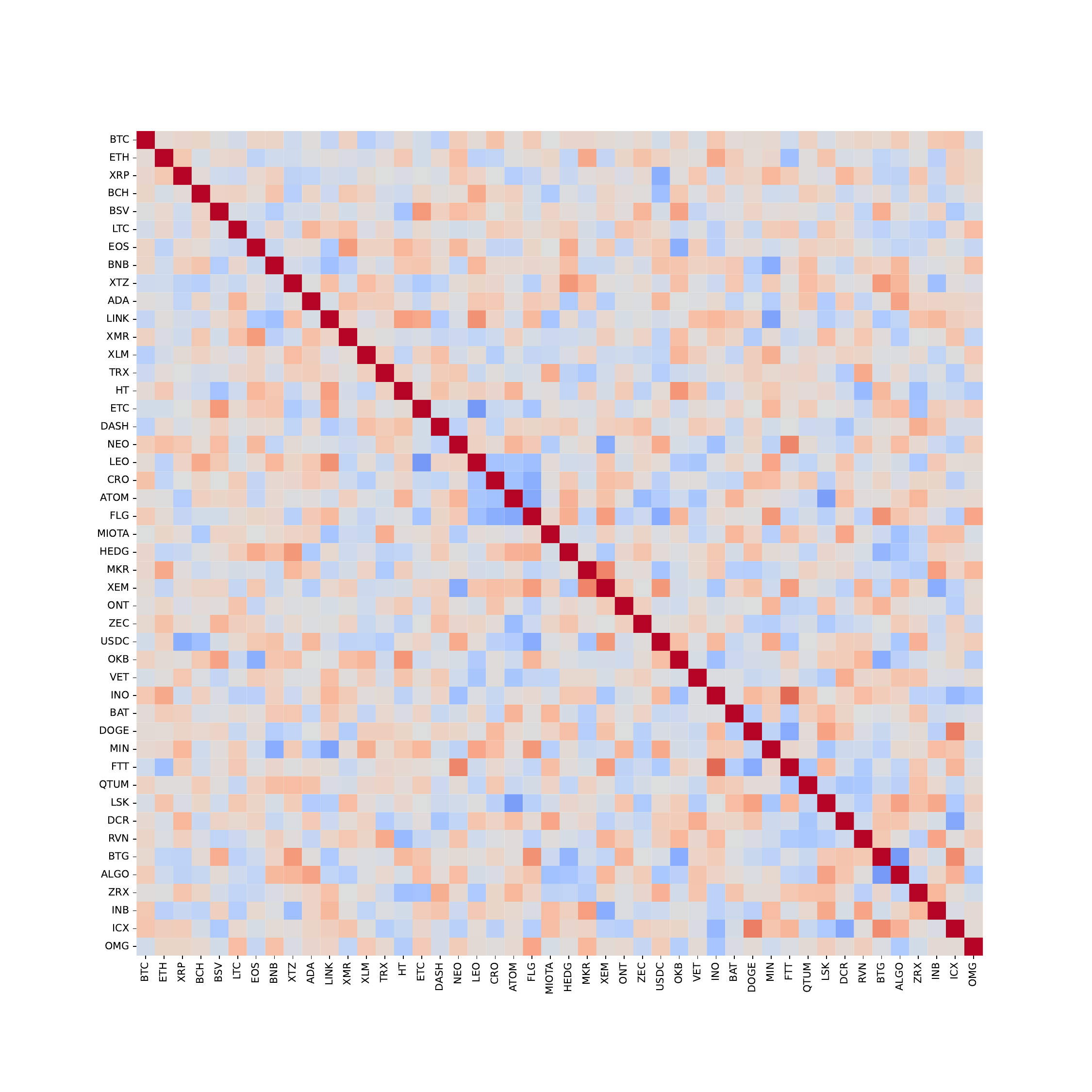}
         \caption{Crash (2019-20)}
         %\label{during19}
     \end{subfigure}
     \hfill
     \begin{subfigure}[b]{0.33\textwidth}
         \centering
         \includegraphics[ width=\textwidth]{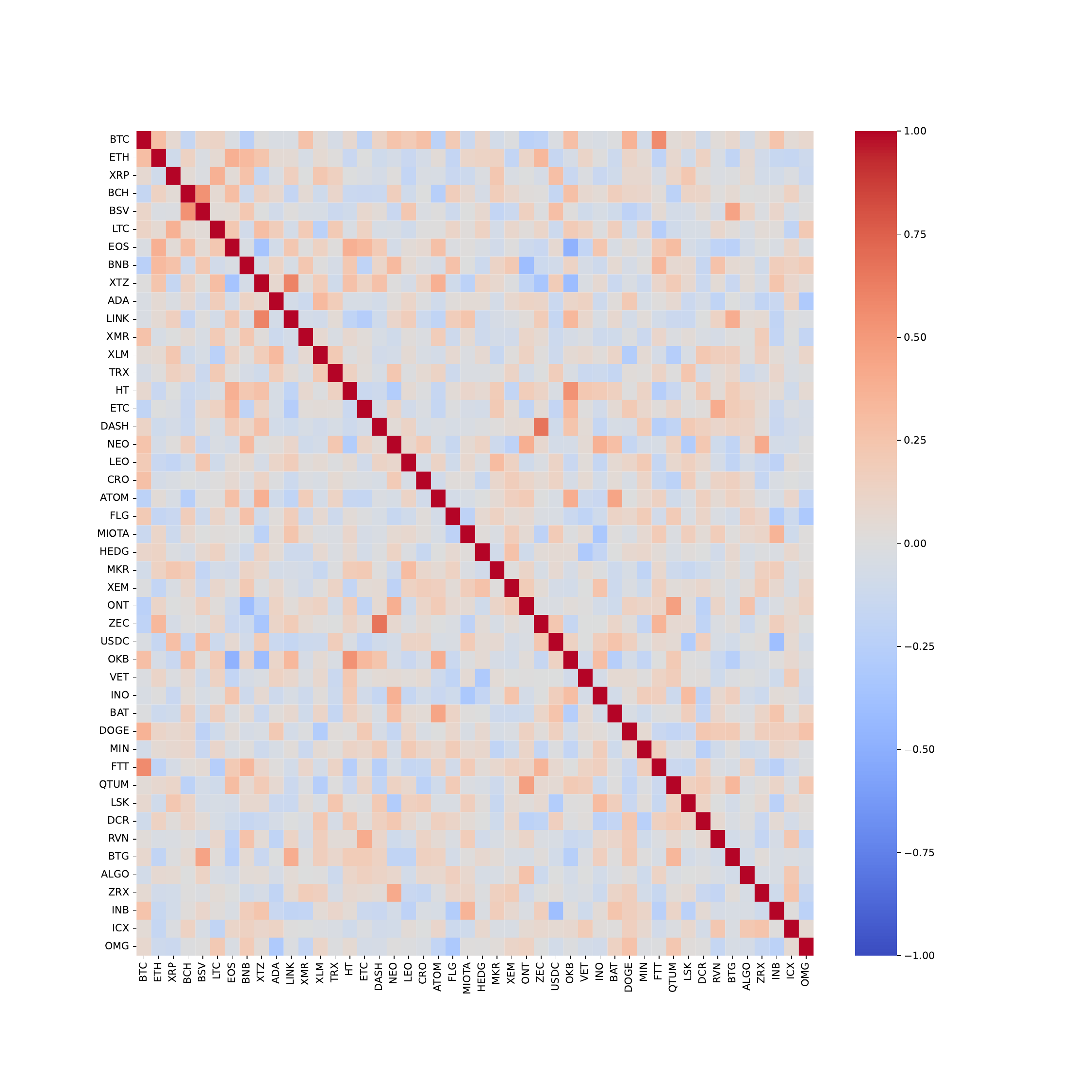}
         \caption{Post-crash (2019-20)}
         %\label{after19}     
     \end{subfigure}
      \caption{The heatmap plot represents the partial correlation matrices among different cryptocurrencies. (a), (b), and (c) represent pre-crash, crash, and post-crash periods for the 2018-19 crash, and (d), (e), and (f) for the 2019-20 crash, respectively.}
        \label{fig:3}
\end{figure}

 We have also obtained similar results for the 2018-19 and 2019-20 crashes,  as shown in Figs.~\ref{fig:3}(a)-(c) and ~\ref{fig:3}(d)-(f) respectively. The mean values of the partial correlations during pre-crash, crash, and post-crash of 2018-19 are 0.304, 0.898, and 0.312, respectively. For the 2019-20 crash, the values are 0.312, 0.339, and 0.316, respectively.

The results indicate that the pre-crash period exhibits the lowest correlation, suggesting a stable market, consistent with findings from studies conducted in Refs.~\cite{nobi2013random,plerou2002random}. The mean value of partial correlation increases significantly during the crash periods, suggesting a synchronized movement of the market during the crash. We find the correlation decreases in the post-crash period. This decrease in the correlation may indicate that the market is returning to a normal period as was in the pre-crash period, ultimately pointing toward a path of stabilization. The transition of correlation values from the pre-crash period to the crash and, subsequently, to the post-crash period captures the dynamics of market behavior at different periods of a crash. 

%\subsection{Network dynamics of cryptocurrencies}
\subsection{Network dynamics of the crashes}
\label{NDA}

Based on the partial correlation among the cryptocurrencies, we define a threshold value (\(\theta\)) as discussed in Sec.~\ref{Threshold Method} and construct a network graph for pre-crash, crash, and post-crash periods for different market crashes. The \(\theta\) plays a crucial role in constructing the network by determining which connections or interactions are significant, thereby shaping the structure and behavior of the network. In this study, the cryptocurrency connections whose partial correlation is greater than the \(\theta\) are considered, and hence, the network size varies for different periods. Each cryptocurrency is represented by the nodes and the edges show their partial correlation which are greater than the \(\theta\). The time intervals for the network construction during the pre-crash, crash, and post-crash periods are shown in Table~\ref{tab:DT}.

\begin{figure}[ht]
     \centering
     \begin{subfigure}[b]{0.32\textwidth}
         \centering
         \includegraphics[width=\textwidth]{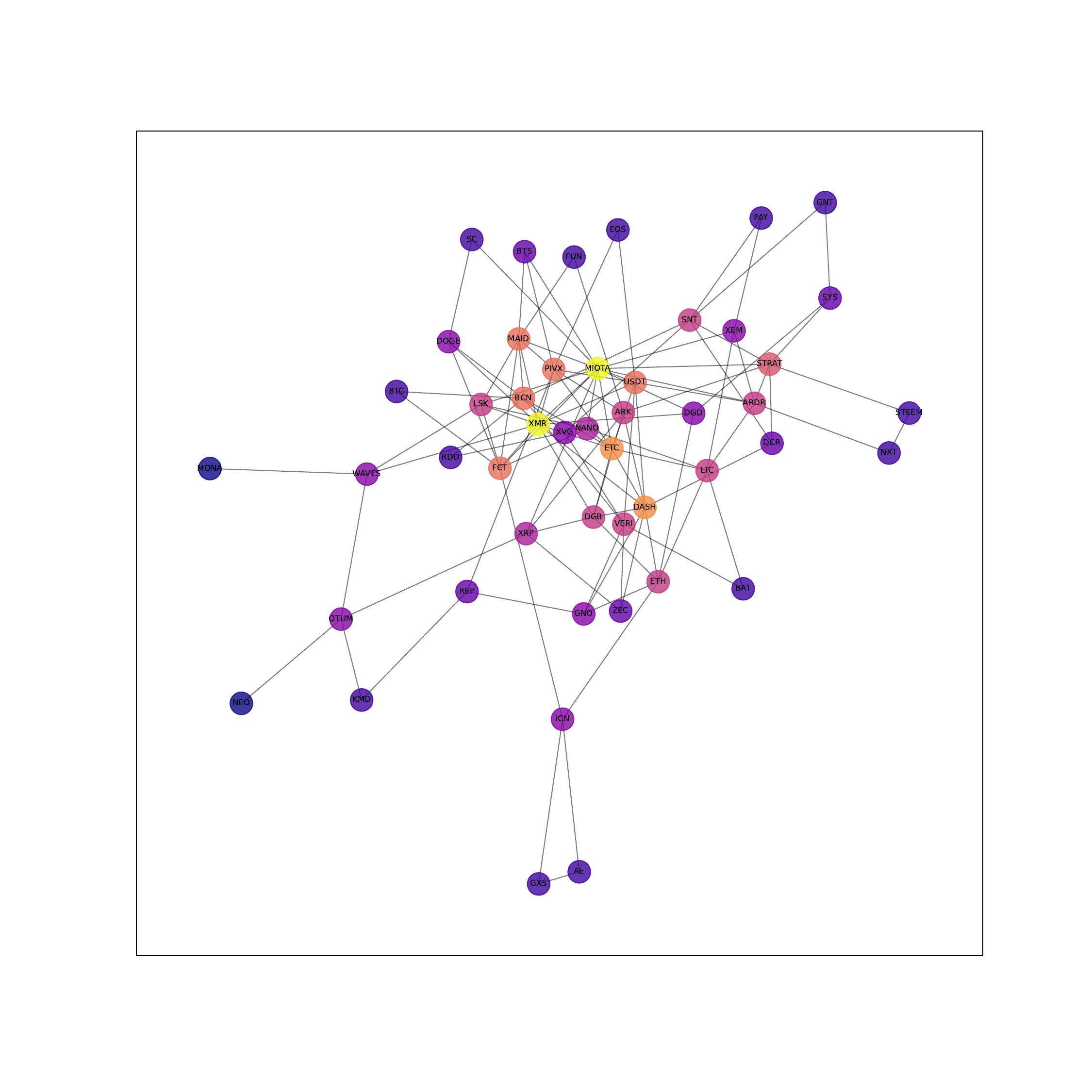}
         \caption{Pre-crash (2017-18)}
         \label{17-18B}
     \end{subfigure}
     \hfill
     \begin{subfigure}[b]{0.33\textwidth}
         \centering
         \includegraphics[width=\textwidth]{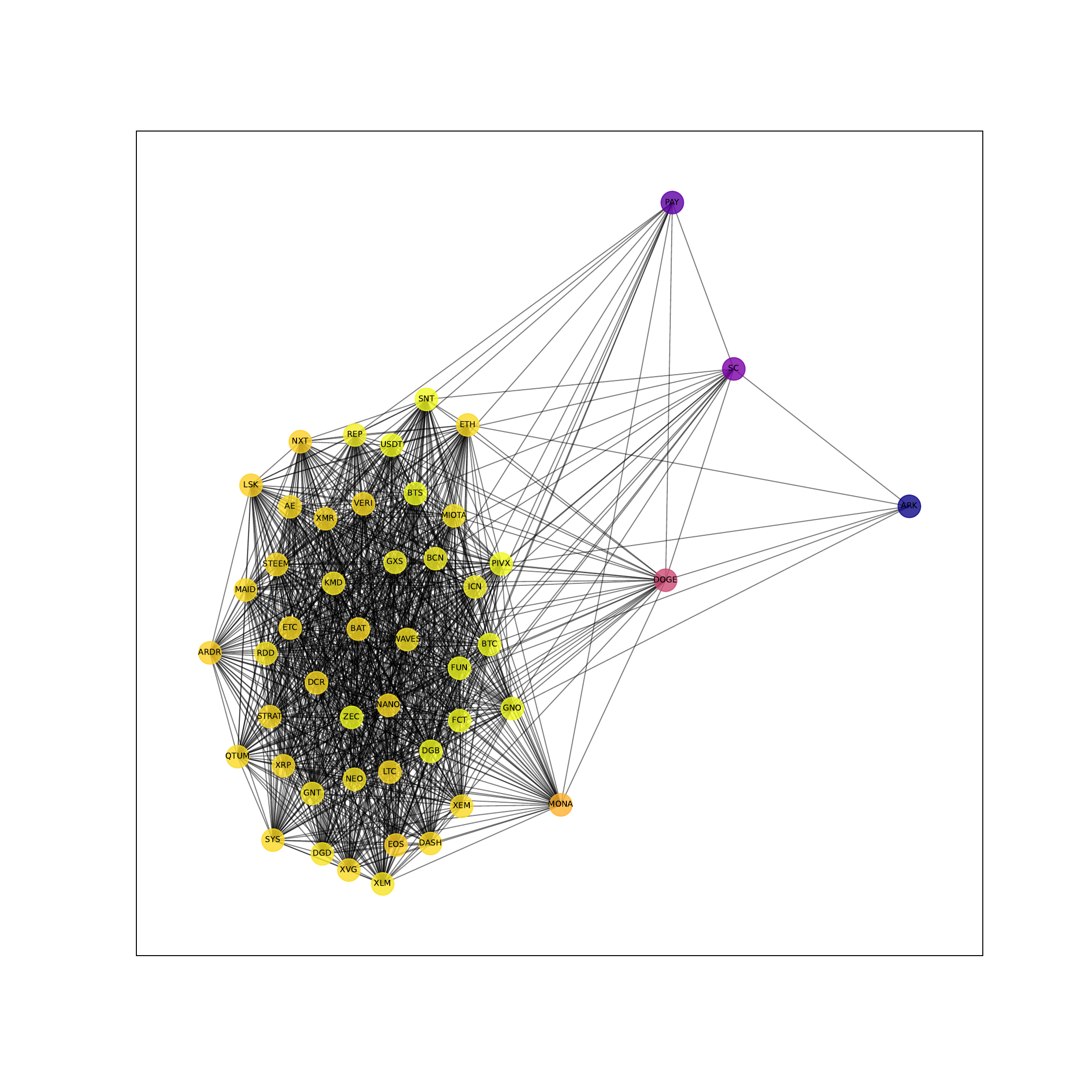}
         \caption{Crash (2017-18)}
         \label{17-18C}
     \end{subfigure}
     \begin{subfigure}[b]{0.33\textwidth}
         \centering
         \includegraphics[width=\textwidth]{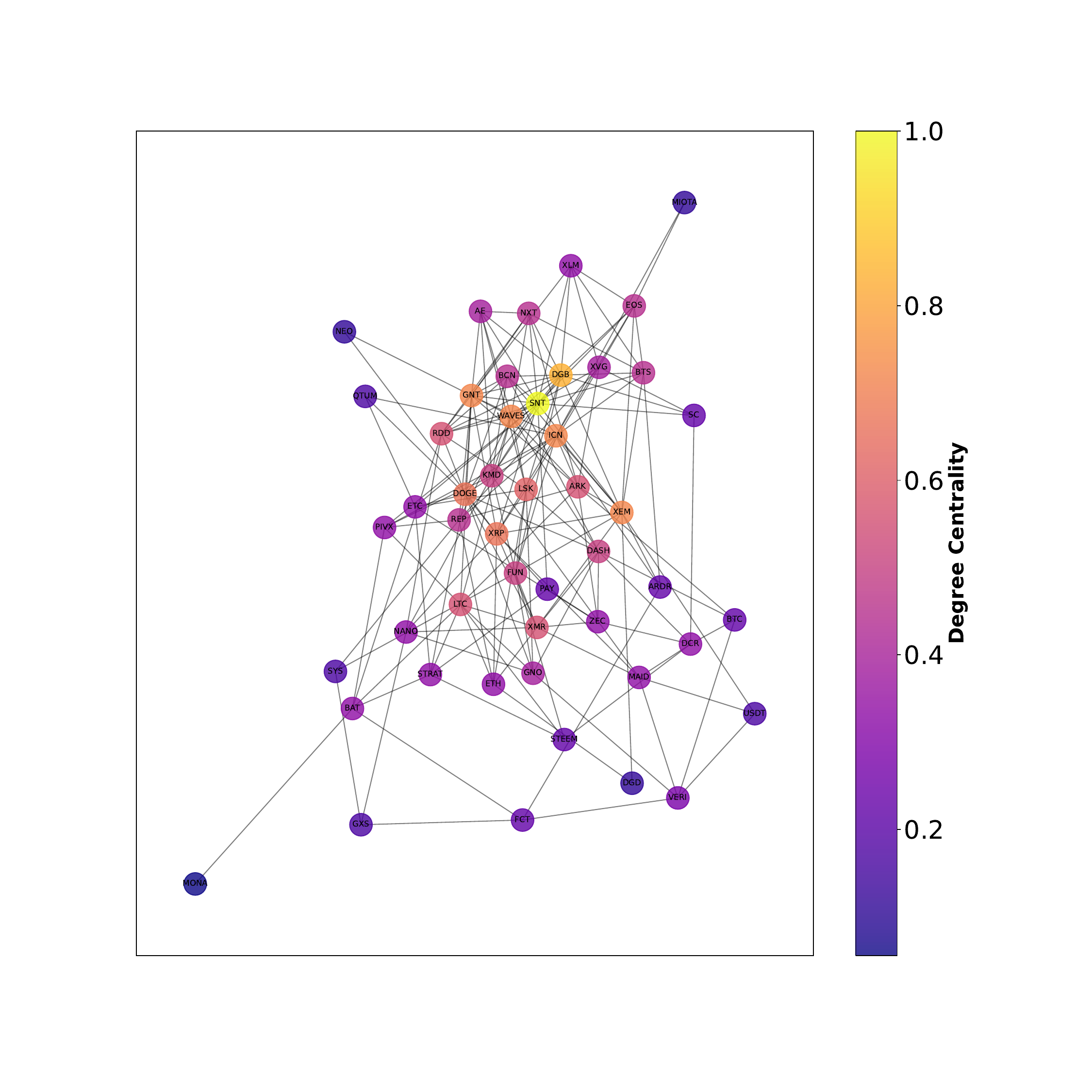}
         \caption{Post-crash (2017-18)}
         \label{17-18A}
     \end{subfigure}
      \caption{The figure represents the network of 2017-18 cryptocurrency crashes during different periods. Plots (a), (b), and (c) represent the network during pre-crash, crash, and post-crash periods for the 2017-18 crash. A dense network is formed during the crash periods.}
        \label{fig:4}
\end{figure}

Figs.~\ref{fig:4}(a),~\ref{fig:4}(b), and~\ref{fig:4}(c) show the network distribution of different cryptocurrencies during the pre-crash, crash, and post-crash periods of the 2017-18 crash, respectively. We observe that in the crash period, the connections between the cryptocurrencies are very dense with a higher connectedness. This indicates a synchronized price movement as a result of uniform reactions from the investors. Such uniform reaction results in herding behavior, often leading to a panic sell-off in the market~\cite{rai2020statistical}. However, in the post-crash period, the connections become less dense as the market enters a period of adjustment, characterized by various market corrections. In the pre-crash period, we observed the least connections, suggesting a stable and normal market environment characterized by independent movements rather than collective shifts.

\begin{figure}[ht]
     \centering
     \begin{subfigure}[b]{0.32\textwidth}
         \centering
         \includegraphics[width=\textwidth]{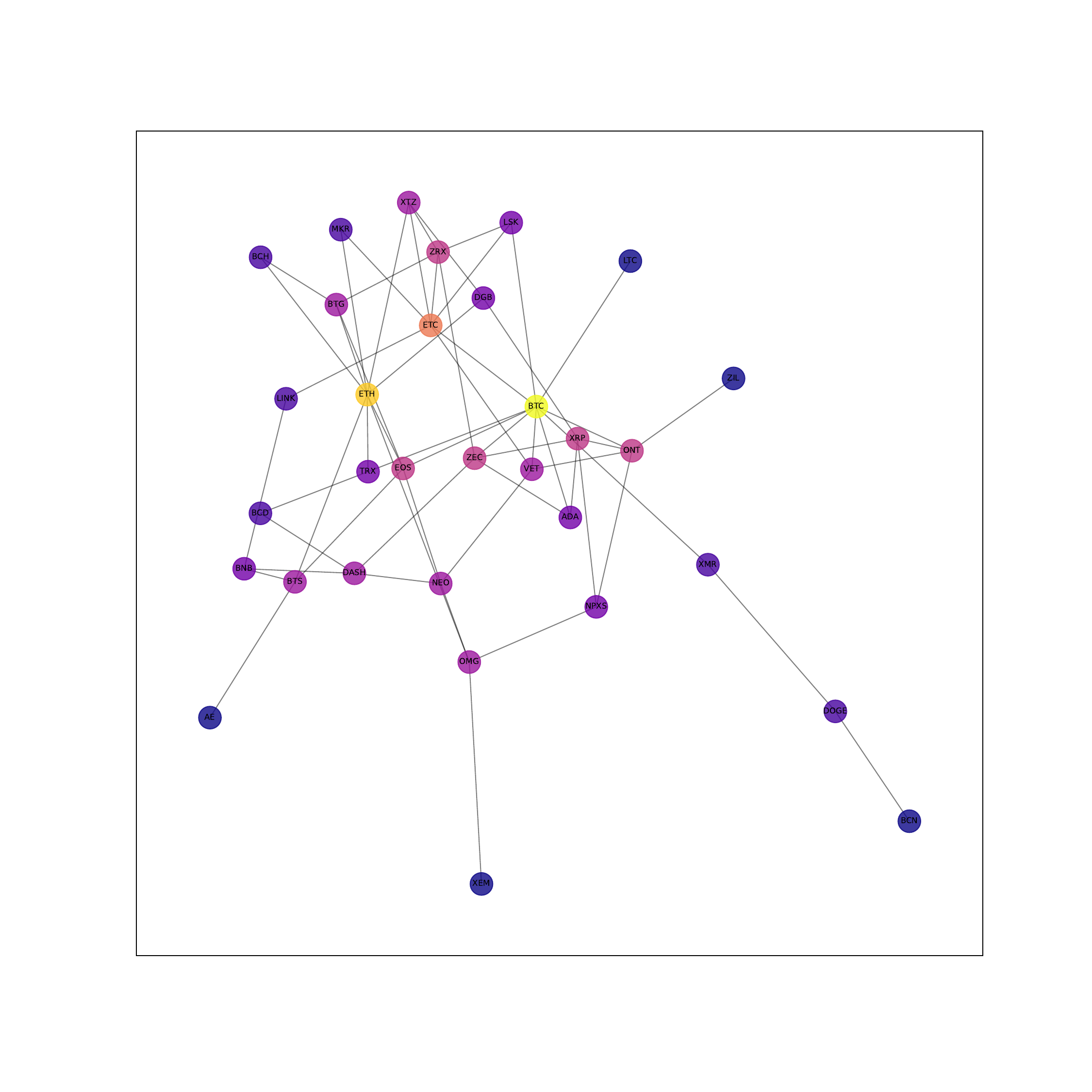}
         \caption{Pre-crash (2018-19)}
         \label{18-19B}
     \end{subfigure}
     \hfill
     \begin{subfigure}[b]{0.33\textwidth}
         \centering
         \includegraphics[width=\textwidth]{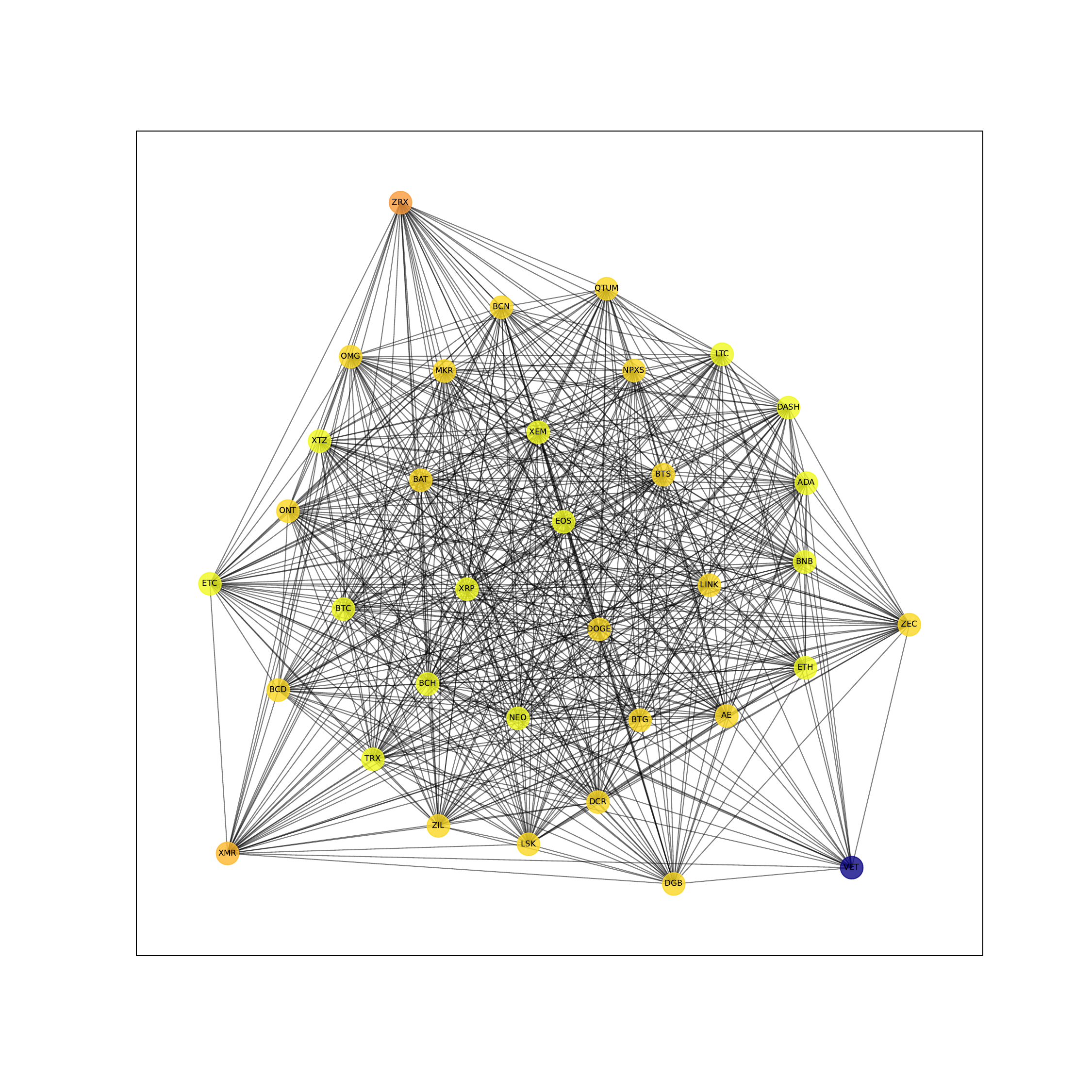}
         \caption{Crash (2018-19)}
         \label{18-19C}
     \end{subfigure}
     \begin{subfigure}[b]{0.33\textwidth}
         \centering
         \includegraphics[width=\textwidth]{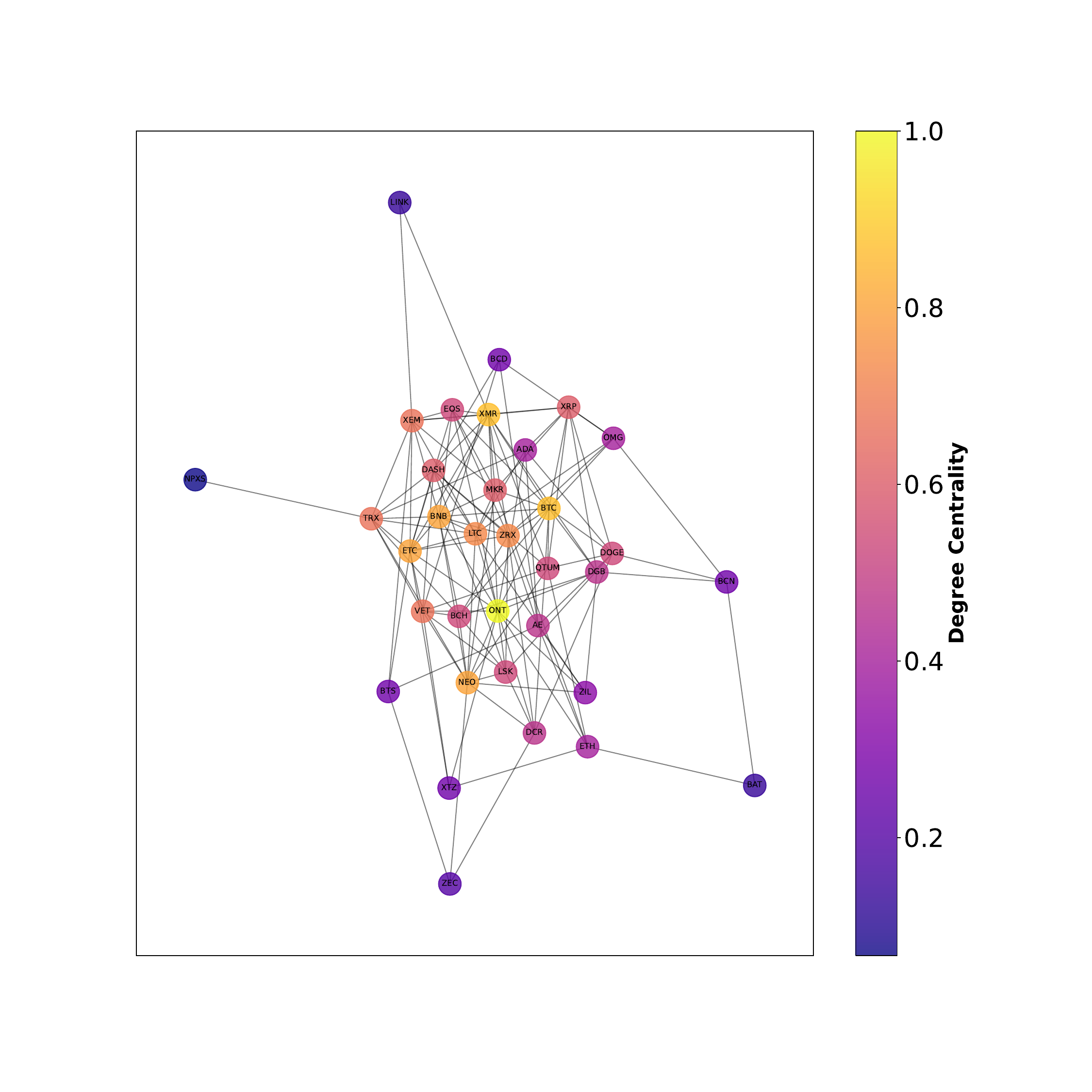}
         \caption{Post-crash (2018-19)}
         \label{18-19A}
     \end{subfigure}
      \caption{The figure represents the network of 2018-19 cryptocurrency crashes during different periods. Plots (a), (b), and (c) represent the network during pre-crash, crash, and post-crash periods for the 2018-19 crash. A dense network is formed during the crash periods.}
        \label{fig:5}
\end{figure}

\begin{figure}[ht]
     \centering
     \begin{subfigure}[b]{0.32\textwidth}
         \centering
         \includegraphics[width=\textwidth]{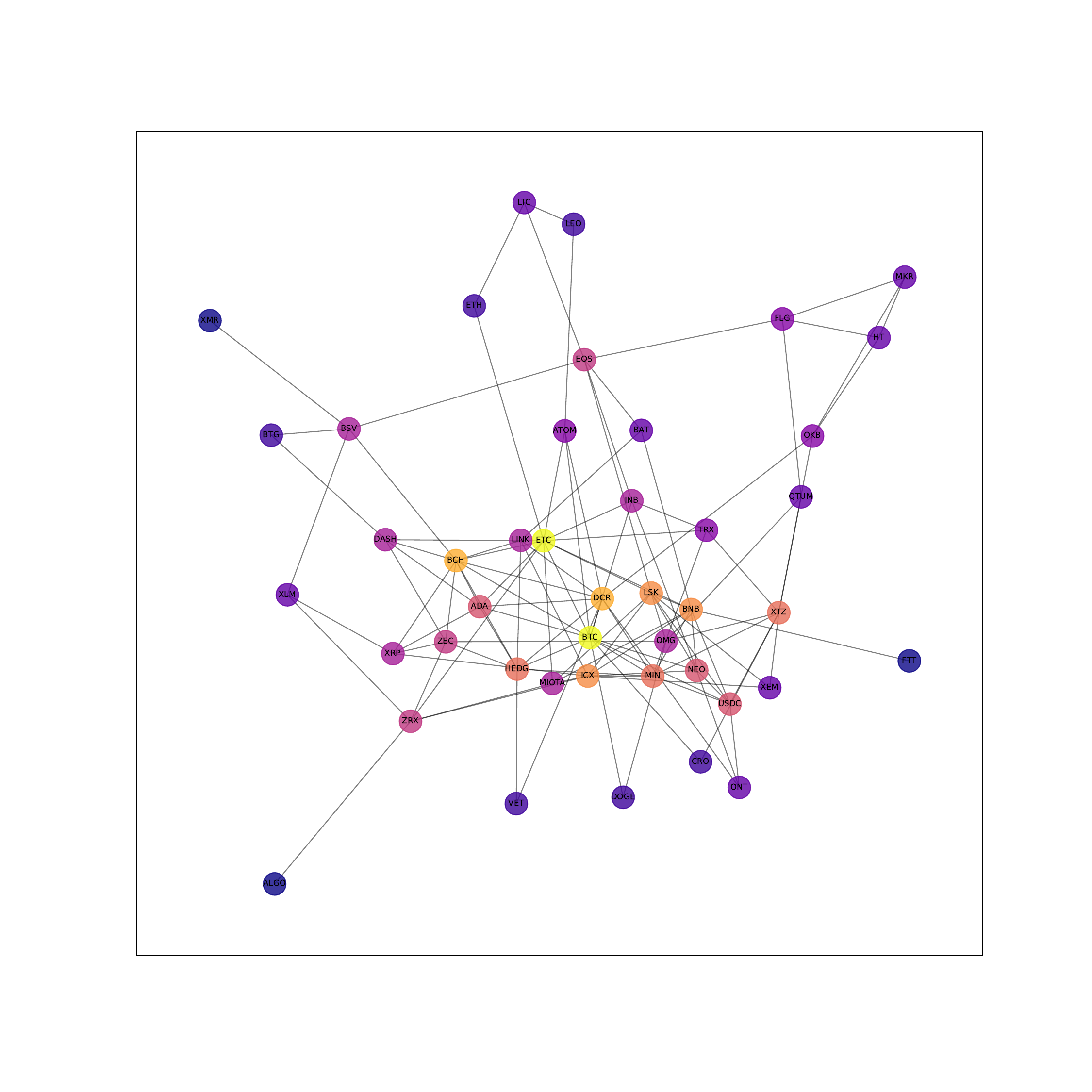}
         \caption{Pre-crash (2019-20)}
         \label{19-20B}
     \end{subfigure}
     \hfill
     \begin{subfigure}[b]{0.33\textwidth}
         \centering
         \includegraphics[width=\textwidth]{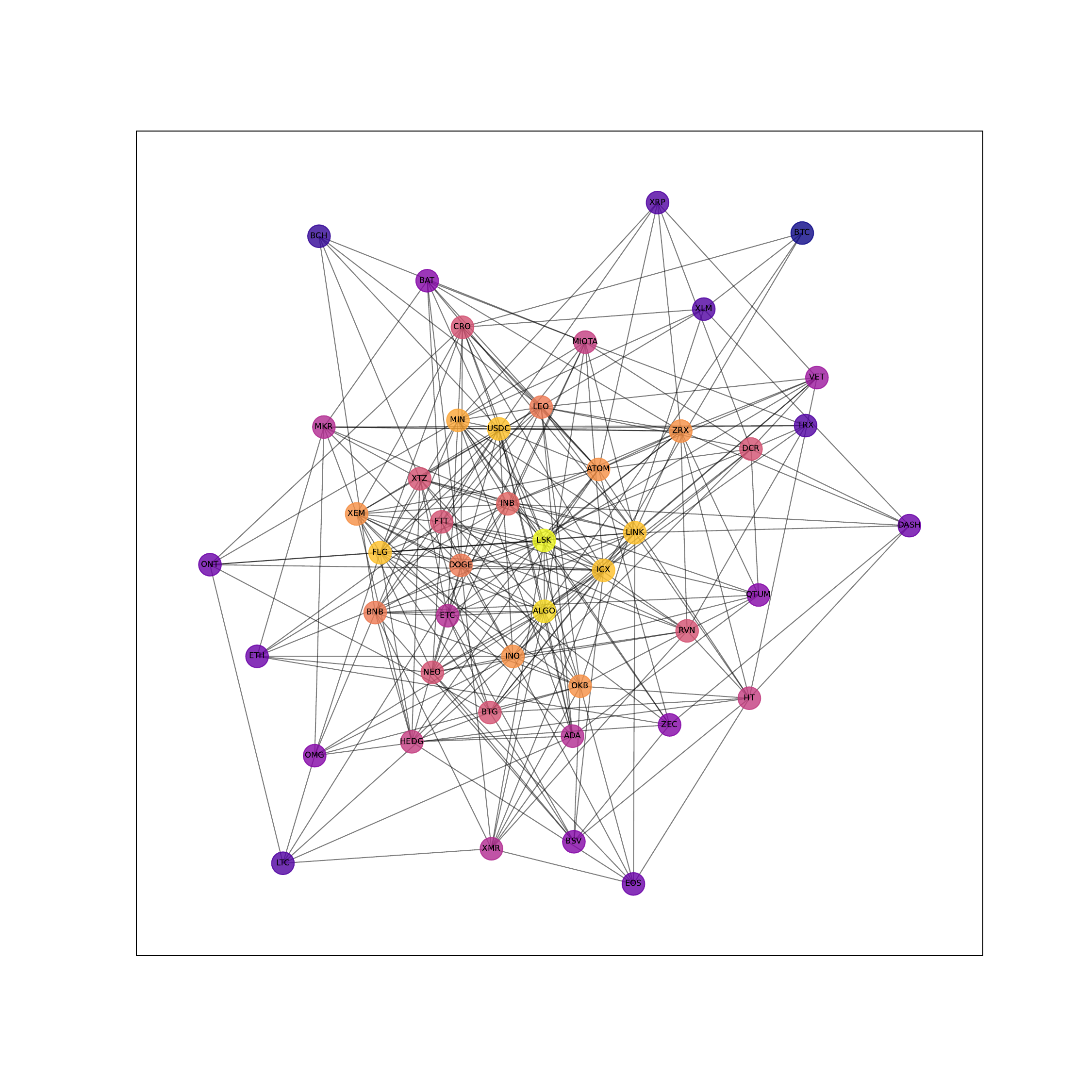}
         \caption{Crash (2019-20)}
         \label{19-20C}
     \end{subfigure}
      \begin{subfigure}[b]{0.33\textwidth}
         \centering
         \includegraphics[width=\textwidth]{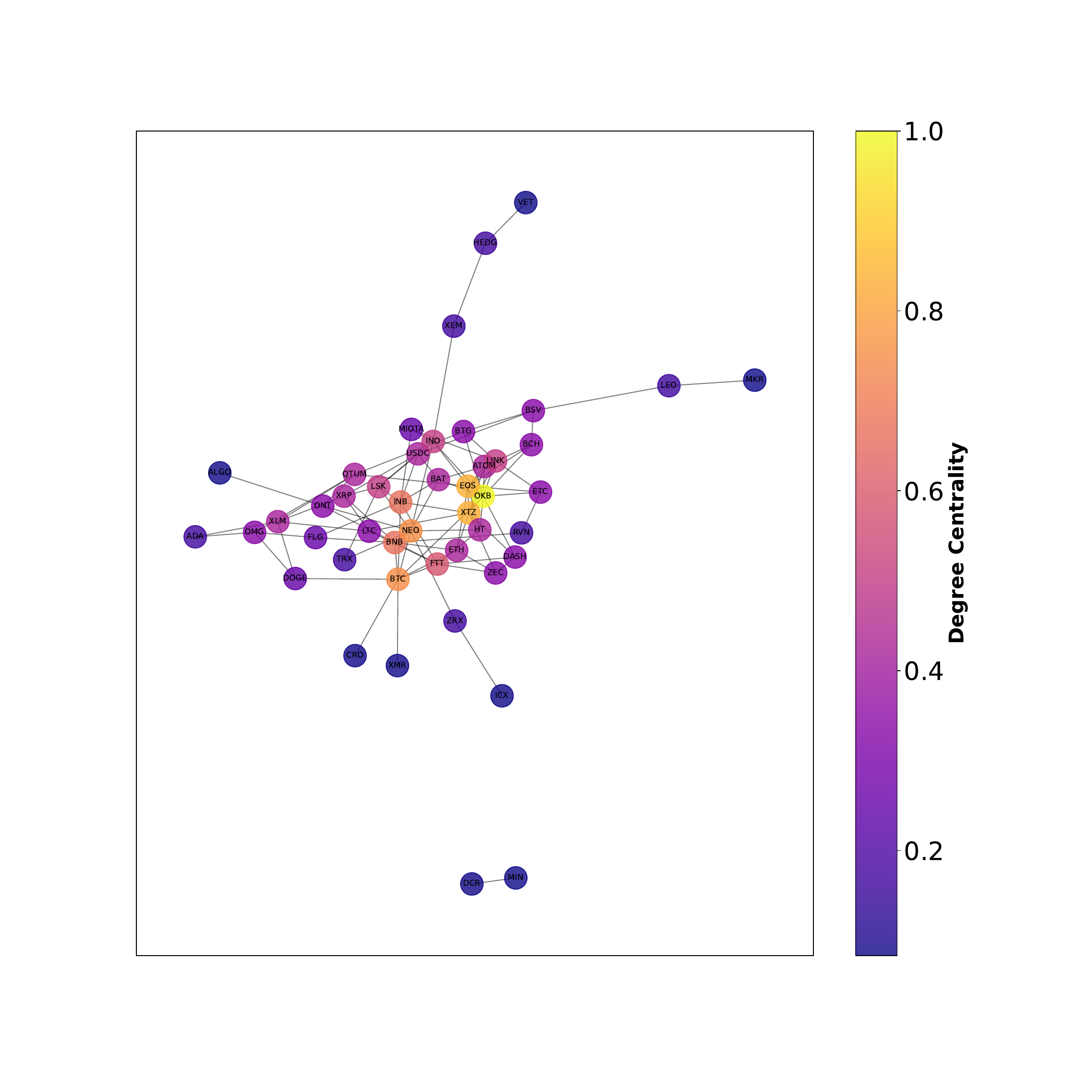}
         \caption{Post-crash (2019-20)}
         \label{19-20A}
     \end{subfigure}
      \caption{The figure represents the network of 2019-20 cryptocurrency crashes during different periods. Plots (a), (b), and (c) represent the network during pre-crash, crash, and post-crash periods for the 2019-20 crash period.}
        \label{fig:6}
\end{figure}

Similar changes between pre-crash, crash, and post-crash periods are observed in 2018-19 crashes, as shown in Figs.~\ref{fig:5}(a)-(c). For the 2019-20 crash, we observed a different pattern where the pre-crash period shows denser connections compared to the post-crash period, as shown in Figs.~\ref{fig:6}(a)-(c). This distinct pattern may be due to the different dynamics leading to the crash. To further analyze and quantify these changes in network dynamics across different periods of these crashes, we have calculated the degree density ($\rho_D$), average path length (\(\bar{l}\)), and average clustering coefficient ($\overline{cc}$).

\begin{table}[ht]
\caption{Table contains the degree density ($\rho_D$) of the cryptocurrency network across different crash periods.}
\centering
\begin{tabular}{|>{\centering\arraybackslash}m{2.5cm}|>{\centering\arraybackslash}m{3cm}|>{\centering\arraybackslash}m{3cm}|>{\centering\arraybackslash}m{3cm}|}
\hline
Market period & Pre-crash  &  Crash   & Post-crash  \\ 
\hline
2017-18 & 0.097&  0.884 & 0.181 \\ 
\hline
2018-19  & 0.115 & 0.97 & 0.235 \\ 
\hline
2019-20  & 0.118 & 0.25 & 0.096 \\ 
\hline
\end{tabular}

\label{tab:DD}
\end{table}

    We have estimated the $\rho_D$, $\overline{cc}$, and $\bar{l}$ for all the networks during the 2017-18, 2018-19, and 2019-20 cryptocurrency crashes. $\rho_D$ measures the average number of connections per node, reflecting the network's density and the potential for widespread influence, and $\overline{cc}$ demonstrates a tendency for nodes to cluster tightly and form interconnected groups. $\bar{l}$ measures the minimal steps required to move from one node to another.

\begin{table}[ht]
\caption{Table contains the average clustering coefficient ($\overline{cc}$) of the cryptocurrency network during different crash periods.}
\centering
\begin{tabular}{|>{\centering\arraybackslash}m{2.5cm}|>{\centering\arraybackslash}m{3cm}|>{\centering\arraybackslash}m{3cm}|>{\centering\arraybackslash}m{3cm}|}
\hline
Market period & Pre-crash &  Crash  & Post-crash \\ 
\hline
2017-18 & 0.226 &  0.956 & 0.282 \\ 
\hline
2018-19  & 0.174 & 0.977  & 0.338 \\ 
\hline
2019-20  & 0.241 & 0.316  & 0.167 \\ 
\hline
\end{tabular}
\label{tab:CC} 
\end{table}

Tables~\ref{tab:DD},~\ref{tab:CC} and~\ref{tab:AL} show the $\rho_D$, $\overline{cc}$, and $\bar{l}$ values during the 2017-18, 2018-19, and 2019-20 cryptocurrency crashes, respectively. From these tables, we observe that the $\rho_D$ and $\overline{cc}$ are significantly high while $\bar{l}$ is very low during the crash period as compared to the pre-crash and post-crash periods. This suggests a higher level of connectivity and potential for information flow among cryptocurrencies, which may contribute to increased market volatility and rapid transmission of price changes during crash periods. Hence, a strong, uninformed, and synchronized panic-driven selling spree among traders happens, leading to a major plunge in the cryptocurrency market. However, in the post-crash period, the values of $\rho_D$ and $\overline{cc}$ decrease whereas the $\bar{l}$ value increases but remains larger than the pre-crash period for the 2017-18 and 2018-19 crashes. This may be due to the fact that during the post-crash period, the network dynamics are in a transitional phase where the market is gradually returning to the previous pre-crash period with some volatile phases. While the crash period has passed, the market still shows traces of its impact, indicating a lingering effect of past events and occasional aftershocks. Therefore, the network distribution during the post-crash period can be seen as a blend of characteristics from both the crash and pre-crash periods.

\begin{table}[ht]
\caption{Table contains the average path length ($\bar{l}$) of the cryptocurrency network during different crash periods.}
\centering
\begin{tabular}{|>{\centering\arraybackslash}m{2.5cm}|>{\centering\arraybackslash}m{3cm}|>{\centering\arraybackslash}m{3cm}|>{\centering\arraybackslash}m{3cm}|}
\hline
Market period & Pre-crash   &  Crash     & Post-crash   \\ 
\hline
2017-18 & 2.792 &  1.116 & 2.236 \\ 
\hline
2018-19  & 2.742 &  1.03 & 1.961 \\ 
\hline
2019-20  & 2.578 & 1.82 & 2.901 \\ 
\hline
\end{tabular}
\label{tab:AL}
\end{table}

\section{Conclusion}
\label{Conclusion}

In this paper, we have identified the crashes in the cryptocurrency market and studied their dynamics using complex network analysis. We have identified three cryptocurrency crashes from the year 2017 to 2020, using Hilbert Spectrum. In order to understand the characteristics of crashes, we have divided each market crash into three periods namely pre-crash, crash, and post-crash period. We have calculated the partial correlation during these periods. The partial correlation is highest during the crash period, followed by the post-crash period, and least in the pre-crash period. The partial correlation is further applied to construct the network between the cryptocurrencies during different periods of the three market crashes.

In this paper, we have identified three crashes in the cryptocurrency market from 2017 to 2020, using Hilbert Spectrum, and studied their dynamics using complex network analysis. In order to understand the characteristics of crashes, we have divided each market crash into three periods namely pre-crash, crash, and post-crash period. We have calculated the partial correlation during these periods. The partial correlation is highest during the crash period, followed by the post-crash period, and least in the pre-crash period. The partial correlation is further applied to construct the network between the cryptocurrencies during different periods of the three market crashes.

We have constructed the network for different periods of the 2017-18, 2018-19, and 2019-20 cryptocurrency crashes. Degree density ($\rho_D$), average clustering coefficient ($\overline{cc}$), and average path length ($\bar{l}$) are estimated from these networks. During the crash periods of 2017-18, 2018-19, and 2019-20, we observed that $\rho_D$ and $\overline{cc}$ were significantly higher, while $\bar{l}$ was notably lower compared to the pre-crash and post-crash periods. For the 2017-18 and 2018-19 crashes, $\rho_D$ and $\overline{cc}$ decreased in the post-crash period from their levels during the crash and reached their lowest points in the pre-crash period. Whereas, the average path length ($\bar{l}$) increased after the crash and peaked in the pre-crash period. However, during the 2019-20 crash, $\rho_D$ and $\overline{cc}$ were highest during the crash period, decreased in the pre-crash period, and were lowest in the post-crash period. $\bar{l}$ exhibited the opposite pattern, being lowest in the pre-crash period and highest in the post-crash period.

A higher value of $\rho_D$ and $\overline{cc}$, along with a smaller value of $\bar{l}$ obtained during the crash period confirms the formation of the dense network. The dense network shows a rapid and efficient spread of information across the network, potentially accelerating market reactions and increasing volatility. As a result, there is an uninformed, synchronized panic sell-off by traders, resulting in extreme price declines in the crash period. For the 2017-18 and 2018-19 crashes, $\rho_D$ is comparatively less during the post-crash periods with a least during the pre-crash periods, as validated by the sparse connections in the networks. The relatively high $\rho_D$ and $\overline{cc}$ observed during the post-crash period compared to the pre-crash period may indicate the market does not abruptly return to a stable form after the main crash. However, the variations in $\rho_D$, $\bar{l}$, and $\overline{cc}$ during the 2019-20 crash are relatively modest across the pre-crash, crash, and post-crash periods, suggesting a less pronounced interaction among cryptocurrencies compared to the more significant shifts observed during the 2017-18 and 2018-19 crashes. 

Understanding the network dynamics during different periods of cryptocurrency market crashes reveals a complex interaction. Expanding this research could enhance our broader understanding, potentially guiding investors and traders toward more informed decisions in their future market activities.

\section{Acknowledgement}
We extend our sincerest gratitude to the institute for the generous fellowship support, and our colleague, Buddhanath Sharma for his invaluable contributions to the writing and publication of our research article.

\bibliographystyle{elsarticle-num} 
\bibliography{cas-refs}

\end{document}